\documentclass[letterpaper, 10 pt, conference]{ieeeconf}  
\IEEEoverridecommandlockouts                              
\usepackage{mathrsfs} 
\usepackage{kpfonts,comment} 
\usepackage{times}   
\usepackage{cancel}
\usepackage{xz_style}
\usepackage{setspace}
\usepackage{booktabs}
\usepackage{tabularx} 
\usepackage{siunitx}
\usepackage{sidecap}
\usepackage{{makecell}}
\setcellgapes{3pt}\makegapedcells
\usepackage{cite}

\newcommand{\vertrule}[1][1.5ex]{\rule{.8pt}{#1}}

\renewcommand\theequation{\arabic{section}.\arabic{equation}}

\renewcommand\thesection{\arabic{section}}

\title{\LARGE \bf 
Policy Optimization for PDE Control with a Warm Start}

\author{Xiangyuan Zhang \quad Saviz Mowlavi \quad Mouhacine Benosman \quad Tamer Ba\c{s}ar
\thanks{X. Zhang and T. Ba\c{s}ar are with the Department of ECE and CSL, University of Illinois Urbana–Champaign, Urbana, IL 61801 USA (e-mails: \{xz7, basar1\}@illinois.edu). S. Mowlavi and M. Benosman are with Mitsubishi Electric Research Laboratories (MERL), Cambridge, MA 02139 USA (e-mails: \{mowlavi, benosman\}@merl.com).}
\thanks{X. Zhang and T. Ba\c{s}ar were supported in part by the US Army Research Laboratory (ARL) Cooperative Agreement W911NF-17-2-0181 and in part by the Army Research Office (ARO) MURI Grant AG285. S. Mowlavi and M. Benosman were supported solely by MERL. This work was partially performed during X. Zhang's internship at MERL.}
}

\begin{document}

\maketitle
\thispagestyle{empty}
\pagestyle{empty}

\begin{abstract}
Dimensionality reduction is crucial for controlling nonlinear partial differential equations (PDE) through a ``reduce-then-design'' strategy, which identifies a reduced-order model and then implements model-based control solutions. However, inaccuracies in the reduced-order modeling can substantially degrade controller performance, especially in PDEs with chaotic behavior. To address this issue, we augment the \emph{reduce-then-design} procedure with a policy optimization (PO) step. The PO step fine-tunes the model-based controller to compensate for the modeling error from dimensionality reduction. This augmentation shifts the overall strategy into \emph{reduce-then-design-then-adapt}, where the model-based controller serves as a warm start for PO. Specifically, we study the state-feedback tracking control of PDEs that aims to align the PDE state with a specific constant target subject to a linear-quadratic cost. Through extensive experiments, we show that a few iterations of PO can significantly improve the model-based controller performance. Our approach offers a cost-effective alternative to PDE control using end-to-end reinforcement learning.
\end{abstract}

\section{Introduction}
Closed-loop control of spatio-temporal systems such as turbulent flows promises enhanced energy efficiency in various applications, including vehicle dynamics, chemical and combustion processes, and heating, ventilation, and air conditioning (HVAC) systems \cite{brunton2015closed}. The primary challenges of controlling such systems are the strong nonlinearity and the infinite dimensionality of the governing nonlinear partial differential equations (PDEs). As a result, computationally permissible control strategies typically involve discretization and dimensionality reduction, followed by applying standard model-based control solutions. This process is known as the \emph{reduce-then-design} approach \cite{atwell2001reduced, hinze2005proper, leibfritz2007numerical, hovland2008explicit, barbagallo2009closed}. To date, numerous research efforts have focused on developing more accurate reduced-order models (ROMs) for various PDEs \cite{peitz2019koopman,lee2020model, bhattacharya2021model, fresca2021comprehensive, ahmed2021closures, kaiser2021data}, thus refining the ``reduce'' part of the two-step process. Although more precise ROMs certainly improve control performance, their increased complexity and nonlinearity can diminish the computational benefits. 

In contrast, we focus here on enhancing the ``design'' part of the process given a coarse ROM. Existing model-based controllers often come with optimality guarantees contingent on the specific dynamical models they are based on \cite{lions1971optimal, anderson1990optimal, bacsar1995h, krstic2008boundary,burns2016feedback}. However, the potentially large inaccuracies from reduced-order modeling, as to trade off computational efficiency, can drastically impair control performance,  especially in PDEs that model chaotic systems such as turbulent flows. To address this issue, we augment the \emph{reduce-then-design} procedure with a model-free policy optimization (PO) step that fine-tunes the model-based control gains to compensate for the coarse modeling. This shifts the overall control strategy into \emph{reduce-then-design-then-adapt}, where the model-based control gains serve as a warm start for model-free PO.

Specifically, we study the state-feedback tracking control problem that aims to align the PDE state with a specific constant target subject to an infinite-horizon linear-quadratic (LQ) cost. First, we discretize the PDE in space and time at a set of grid points to arrive at a finite but high-dimensional nonlinear system. Then, we apply the off-the-shelf Dynamic Mode Decomposition with control (DMDc) \cite{proctor2016dynamic} to compute a linear surrogate model that reduces the state dimension by at least tenfold. Lastly, we compute the model-based LQ tracking controller and apply it as a warm start for model-free PO. The PO step further fine-tunes the controller gains using nonlinear high-dimensional PDE solutions until convergence.

We demonstrate the effectiveness of the additional PO step through extensive experiments on three nonlinear PDE control tasks governed by Burgers', Allen-Cahn, and Korteweg-de Vries equations, respectively. With a thirty-two-fold dimensionality reduction in modeling, model-free PO reduces the cost of the model-based LQ tracking controller by $28.0\%$, $15.8\%$, and $36.4\%$, respectively, after only a few iterations. Furthermore, we show that the warm start significantly accelerates the PO process and leads to a more stable training process toward convergence. Our proposed strategy offers a cost-effective solution to PDE control that combines the strengths of model-based and data-driven approaches. It optimizes the \emph{reduce-then-design} process in applications where fine-grained modeling is impractical. As a by-product, the concept of a computationally cheap warm start could also be beneficial in applying end-to-end reinforcement learning (RL) to PDE control.

\subsection{Related Literature}\label{sec:literature}

Most PDEs produce solutions that reside on a manifold of moderate dimension in the phase space \cite{cohen2015approximation}. ROMs take advantage of this property by discovering an approximation of this manifold and a dynamical model within \cite{benner2015survey, rowley2017model}. Decades of research have resulted in numerous data-driven techniques to construct linear and nonlinear ROMs \cite{holmes2012turbulence, brunton2022data, kramer2024learning}. We employ DMDc due to its offline nature, ability to model the control mapping, and that the generated linear surrogate model enables closed-form computations of optimal controllers. Our proposed strategy generalizes to more complex ROMs, controller parametrizations, and PO schemes.

Our work bridges model-based and data-driven approaches for closed-loop PDE control. Traditionally, ROM-based controllers have been proposed following the \emph{reduce-then-design} approach \cite{sipp2016linear, gao2017active, tsolovikos2020estimation}, but their control performance has been limited by the accuracy of the ROM. To overcome this limitation, a recent trend is to directly learn controllers from data, assuming access to a simulator of the real system. These approaches include genetic algorithms \cite{duriez2017machine}, Bayesian optimization \cite{blanchard2021bayesian}, and deep RL \cite{fan2020reinforcement, Garnieretal21}, but they require formidable computational or experimental resources.

Lastly, our work is relevant to the literature on PO for control \cite{vrabie2009adaptive, schulman2015high,lillicrap2015continuous,recht2019tour, zhang2023controlgym} and the theoretical foundation underlying this approach \cite{fazel2018global, zhang2019policymixed, qu2020combining, zhang2021derivative, zhang2023revisiting, zhang2023learning, hu2023toward, zhang2023ifac, zhang2023global}. In particular, \cite{qu2020combining} studied the convergence of PO in linear systems with a ``small'' nonlinear perturbation, where the linear dynamics were utilized to warm start PO for adapting to the nonlinear perturbation. In contrast to \cite{qu2020combining}, we study general nonlinear PDEs where an explicit linear component is not readily identifiable. We propose the generic strategy of first designing a ROM-based controller and then using data-driven PO for fine-tuning.

\section{Problem Formulation}\label{sec:formulation}
We consider one-dimensional PDE control problems with periodic boundary conditions and spatially distributed controls, following the definitions in Section 2.2 of \cite{zhang2023controlgym}. The spatial domain is defined as $\Omega = [0, L] \subset \mathbb{R}$, where $L$ is its length, and $u(x, t):\Omega \times \mathbb{R}^+ \rightarrow \mathbb{R}$ represents a continuous field over spatial and temporal coordinates $x$ and $t$, respectively. The PDE dynamics is described by:
\begin{align}\label{eqn:pde}
	\frac{\partial u}{\partial t} - \mathcal{F} \left( \frac{\partial u}{\partial x}, \frac{\partial^2 u}{\partial x^2}, \dots \right) =  a(x, t),
\end{align}
with $\mathcal{F}$ being a nonlinear differential operator involving spatial derivatives of various orders and depending on various physical parameters, and $a$ representing a distributed control as $a(x, t) = \sum_{j = 0}^{n_a-1} \Phi_j(x) a_j(t)$. Here, $a(x, t)$ integrates $n_a$ scalar inputs $a_j(t)$, each modulated by its forcing support function $\Phi_j(x)$, to model energy addition or external influences on the PDE dynamics. The periodic boundary conditions enforce the continuity of $u$ and its spatial derivatives at the boundary of $\Omega$. We use  $u_r(x)$ to denote the constant target state with which we aim to align the field $u(x, t)$.

We discretize \eqref{eqn:pde} in space and time, defining the high-
dimensional state $z_k \in \RR^{n_z}$ to represent $u$ at $n_z$ equally-spaced points in $\Omega$ at discrete time $k \in \NN$, with $t = k \, \Delta t$ and $\Delta t \in \RR^+$ as the user-defined sampling time. We treat scalar control inputs $a_i(t)$ as piecewise constant over each time step $\Delta t$. The PDE dynamics are then approximated by
\begin{align}\label{eqn:nonlinear_discrete}
	z_{k+1} = f(z_k, a_k),
\end{align}
where $f:\RR^{n_z}\times\RR^{n_a}\to\RR^{n_z}$ is a time-invariant function contingent on the physical parameters and forcing support functions $\Phi_i$ of the specific PDE, and the initial condition $z_0$ is sampled from a distribution $\cD$.

\subsection{State-Feedback Tracking Control}
Our objective is to design a state-feedback controller that enables asymptotic tracking of the discretized constant target state $z_{r} \in \RR^{n_z}$. Specifically, we aim to minimize the LQ tracking cost defined as
\begin{align}\label{eqn:LQT_objective}
	&\hspace{1.6em} J := \EE_{z_0}\bigg\{\sum_{k=0}^{\infty} (z_k-z_{r})^{\top}Q(z_k-z_{r}) + a_k^{\top}Ra_k\bigg\} \\
	&\text{s.t.} \ z_{k+1} = f(z_k, a_k), \ a_k = \phi(z_0, \cdots, z_k, a_0, \cdots, a_{k-1}), \nonumber
\end{align}
where $Q \geq 0$ and $R > 0$ are the symmetric positive (semi)-definite weighting matrices. The control policy $\phi$ maps all available information up to $k$ to the control input $a_k$. Setting $z_{r}$ to $0$ simplifies our setting to the LQ regulation task. Due to the nonlinearity of $f$, there are no general closed-form solutions for the optimal $\phi$ that minimizes the cost  \eqref{eqn:LQT_objective}.

\subsection{Reduced-Order Model Identification}
To address the computational challenges posed by a high-dimensional state $z_k \in \RR^{n_z}$, we adopt a two-step strategy: first performing a dimensionality reduction and then designing a state-feedback controller using the ROM. Among numerous data-driven methods to construct a ROM \cite{taira2017modal}, we utilize the DMDc algorithm \cite{proctor2016dynamic} described in Algorithm \ref{alg:DMDc}. 

DMDc generates a reduced-order linear model $(A, B)$ using $N$ consecutive snapshots from a single trajectory of the nonlinear system \eqref{eqn:nonlinear_discrete}. The dimensionality reduction is defined by an orthogonal mode-spanning matrix $U \in \RR^{n_z \times n_s}$ generated from DMDc such that $z_k \approx Us_k$, with $s_k \in \RR^{n_s}$ denoting the reduced-order state with dimension $n_s \ll n_z$. Then, DMDc identifies the best-fit linear ROM described by
\begin{align}\label{eqn:linear_discrete}
	s_{k+1} = As_k + Ba_k,
\end{align}
where $s_0$ is the projection of the initial condition $z_0$ onto the reduced-order space, given by $s_0 = U^{\top}z_0$. 

\begin{figure}\vspace{-0.5em}
\end{figure}
\begin{algorithm}[t]
\caption{DMDc \cite{proctor2016dynamic}}\label{alg:DMDc}
\begin{algorithmic}[1]
\renewcommand{\algorithmicrequire}{\textbf{Input:}}
 \renewcommand{\algorithmicensure}{\textbf{Output:}}
 \REQUIRE A single trajectory of \eqref{eqn:nonlinear_discrete}, i.e., $\{z_0, \cdots, z_{N}\}$ and $\{a_0, \cdots, a_{N-1}\}$, truncation values $p$ and $n_s$.
\ENSURE Reduced-order system matrices $A \in \RR^{n_s\times n_s}$, $B \in \RR^{n_s\times n_a}$, 
					and the projection matrix $U \in \RR^{n_z\times n_s}$.
\STATE Construct the data matrix \small{$\Omega = \begin{bmatrix}
	Z \\ \Upsilon
\end{bmatrix}$}, where:

\hspace{-1em}{\footnotesize
	$Z = \left[\arraycolsep=0.8pt\def\arraystretch{0.05}\begin{array}{cccc}
    \vertrule{} & \vertrule{} &        & \vertrule{} \\
    z_{0}    & z_{1}    & \ldots & z_{N-1}    \\
    \vertrule{} & \vertrule{} &        & \vertrule{} 
  \end{array}\right]$, ~$Z' = \left[\arraycolsep=0.8pt\def\arraystretch{0.05}\begin{array}{cccc}
    \vertrule{} & \vertrule{} &        & \vertrule{} \\
    z_{1}    & z_{2}    & \ldots & z_{N}    \\
    \vertrule{} & \vertrule{} &        & \vertrule{} 
  \end{array}\right]$, ~$\Upsilon = \left[\arraycolsep=0.8pt\def\arraystretch{0.05}\begin{array}{cccc}
    \vertrule{} & \vertrule{} &        & \vertrule{} \\
    a_{0}    & a_{1}    & \ldots & a_{N-1}    \\
    \vertrule{} & \vertrule{} &        & \vertrule{} 
  \end{array}\right] $};

\STATE Find the $p$-truncated SVD of $\Omega \approx \Psi\Xi \Lambda^{\top} = {\footnotesize\begin{bmatrix}
	\Psi _{Z} \\ \Psi _{\Upsilon}
\end{bmatrix}}\Xi \Lambda^{\top}$\;
\STATE Compute the $n_s$-truncated SVD of $Z' \approx U\Gamma V^{\top}$\;
\STATE Generate $A = U^{\top}Z'\Lambda\Xi^{-1}\Psi^{\top}_{Z}U$, \ $B = U^{\top}Z'\Lambda\Xi^{-1}\Psi^{\top}_{\Upsilon}$\;

\STATE \textbf{return} $A$, $B$, and $U$.
\end{algorithmic}
\end{algorithm}

\section{Policy Optimization with a Warm Start}
We first formulate the LQ tracking problem based on the ROM \eqref{eqn:linear_discrete} and present its closed-form solution. Specifically, the tracking objective is defined as
\begin{align}\label{eqn:LQTr_objective}
	&\hspace{2em} J_R :=  \EE_{s_0}\bigg\{\sum_{k=0}^{\infty} (s_k-s_{r})^{\top}\tilde{Q}(s_k-s_{r}) + a_k^{\top}Ra_k\bigg\} \\
	&\text{s.t.} \quad s_{r} = U^{\top}z_{r}, \ \tilde{Q} := U^{\top}QU \geq 0, \ s_{k+1} = As_k + Ba_k, \nonumber \\
	&\hspace{2.2em}  a_k = \varphi(s_0, \cdots, s_k, a_0, \cdots, a_{k-1}). \nonumber
\end{align}
The linearity and the low dimensionality of the ROM enable the closed-form computation of the $J_R$-minimizing control policy $\varphi$. Then, it is natural to apply $\varphi$ back to addressing the original high-dimensional problem \eqref{eqn:LQT_objective}. This could be done by projecting $\{z_k\}_{k\geq 0}$ onto the reduced-order space $\RR^{n_s}$ and using the policy $a_k = \varphi(U^{\top}z_0, \cdots, U^{\top}z_k, a_0, \cdots, a_{k-1})$ to generate control inputs. However, dimensionality reduction may introduce significant modeling errors (e.g., when the reduced state dimension $n_s$ is too low or the PDE dynamics are highly nonlinear), which limit the performance of $J_R$-minimizing control policy $\varphi$ when applied to \eqref{eqn:LQT_objective}. Nonetheless, $\varphi$ provides a valuable, computationally cheap starting point for further policy fine-tuning using trajectories from the nonlinear high-dimensional system \eqref{eqn:nonlinear_discrete}. The structure of this section is as follows: Section \ref{sec:LQT} describes the model-based solution to the reduced-order LQ tracking problem \eqref{eqn:LQTr_objective}, which serves as a warm start to the PO algorithm in Section \ref{sec:po} for further fine-tuning. 

\subsection{Model-Based LQ Tracking Controller}\label{sec:LQT}
For the reduced-order LQ tracking problem \eqref{eqn:LQTr_objective}, the optimal controller has the form of \cite{anderson1990optimal}:
\begin{align}\label{eqn:tracker1}
	a_k &= \hspace{-0.1em}-(R+B^{\top}PB)^{-1}B^{\top}PAs_k \hspace{-0.1em}+\hspace{-0.1em} (R+B^{\top}PB)^{-1}B^{\top}q_{k+1}, \\
	q_k &= (A-B(R+B^{\top}PB)^{-1}B^{\top}PA)^{\top}q_{k+1} + \tilde{Q}s_{r}, \quad q_{\infty} = \tilde{Q}s_{r}, \nonumber
\end{align}
where $P$ is the solution of the algebraic Riccati equation
\begin{align*}
	P = A^{\top}PA - A^{\top}PB(R+B^{\top}PB)^{-1}B^{\top}PA + \tilde{Q}, \ \tilde{Q} = U^{\top}QU.
\end{align*}
Moreover, the closed-loop matrix $A-B(R+B^{\top}PB)^{-1}B^{\top}PA$ has a spectral radius less than $1$. Note that for any square matrix $X$ with a spectral radius less than $1$, it holds that $\sum_{k=0}^{\infty}X^k = (I - X)^{-1}$. Therefore, we can rewrite \eqref{eqn:tracker1} as $a_k = K^{MB}_as_k + K^{MB}_bs_{r}$, where $K^{MB}_a, K^{MB}_b$ are the optimal gain matrices independent of the state $s_k$ and the target $s_{r}$:
\begin{align}\label{eqn:tracker_policy1}
 	K^{MB}_a &= -(R+B^{\top}PB)^{-1}B^{\top}PA, \\
 	K^{MB}_b &= (R+B^{\top}PB)^{-1}B^{\top}(I-(A+BK_a^{MB})^{\top})^{-1} \tilde{Q}. \label{eqn:tracker_policy2}
\end{align}
A direct application of the $K^{MB}_a, K^{MB}_b$ to the high-dimensional problem \eqref{eqn:LQT_objective} results in the controller
\begin{align}\label{eqn:reduced_order_controller}
	a_k = K^{MB}_a(U^{\top}z_k) + K^{MB}_b(U^{\top}z_{r}).
\end{align}
When the DMDc model is accurate such that the modeling gap is sufficiently small to be negligible, then \eqref{eqn:reduced_order_controller} is a good candidate for addressing the original problem \eqref{eqn:LQT_objective}, providing desired performance at very low computational cost. However, in the presence of a large modeling gap, the performance of \eqref{eqn:reduced_order_controller} could degrade substantially, which motivates further fine-tuning using PO. 

\subsection{Policy Optimization with a Warm Start}\label{sec:po}
We propose to fine-tune $K^{MB}_a, K^{MB}_b$ using PO to achieve a good balance between performance and computational efficiency, utilizing simulated trajectories of the nonlinear system \eqref{eqn:nonlinear_discrete} until reaching a (local) minima of \eqref{eqn:LQT_objective}. Specifically, we define the PO problem with respect to the concatenated control policy $\pi: = [K_a \ K_b] \in \RR^{n_a \times (2n_s)}$ as
\begin{align}\label{eqn:po}
\hspace{-0.3em}\min_{\pi} \ J(\pi) \quad \text{s.t.} \quad z_{k+1} = f(z_k, a_k), ~ a_k = \pi \begin{bmatrix}
 	U^{\top}z_k \\ U^{\top}z_{r}
 \end{bmatrix},
\end{align}
where the objective function $J(\pi)$ follows \eqref{eqn:LQT_objective}. 

To address \eqref{eqn:po}, we employ a derivative-free policy gradient (PG) method utilizing a (two-point) zeroth-order oracle \cite{flaxman2005online, duchi2015optimal, nesterov2017random}. We define the vanilla PG update rule as
\begin{align}\label{eqn:PG}
	\pi_{i+1} = \pi_i - \eta\cdot \nabla_{\pi_i} J(\pi_i), \quad \pi_0 = [K_a^{MB} \ K_b^{MB}],
\end{align}
where $\eta>0$ is the learning rate and $\nabla_{\pi_i} J(\pi_i)$ is the noisy PG sampled from Algorithm \ref{alg:zeroth-order}.

Upon converging to a local minimum of \eqref{eqn:LQT_objective}, the fine-tuned policy $\pi$ and the dimensionality reduction matrix $U$ are applied in tandem to address the high-dimensional control problem defined in \eqref{eqn:LQT_objective}. The effectiveness and efficiency of the proposed framework will be showcased through extensive numerical experiments in the following section.

\begin{figure}\vspace{-0.5em}
\end{figure}
\begin{algorithm}[t]
\caption{Zeroth-Order Gradient Oracle}\label{alg:zeroth-order}
\begin{algorithmic}[1]
\renewcommand{\algorithmicrequire}{\textbf{Input:}}
 \renewcommand{\algorithmicensure}{\textbf{Output:}}
 \REQUIRE Policy $\pi$, smoothing radius $r$.
\ENSURE The approximated PG $\nabla_{\pi} J(\pi)$.
\STATE Sample a random matrix $\Theta \in \RR^{n_a \times (2n_s)}$ from a zero-mean Gaussian distribution, and normalize it to $\|\Theta\|_F = 1$ \;
\STATE Perform symmetric perturbation to $\pi$ by setting $\pi^{+} = \pi + r\cdot \Theta$ and $\pi^{-} = \pi - r\cdot \Theta$ \;
\STATE Sample $z_{0} \sim \cD$ and perform two rollouts with policy $\pi^+$ and $\pi^-$, respectively. Collect instantiations of the objective value $J(\pi^+)$ and $J(\pi^-)$\;
\STATE \textbf{return} $\nabla_{\pi} J(\pi) = \frac{n_a n_s}{r}[J(\pi^+) - J(\pi^-)]\Theta$\;
\end{algorithmic}
\end{algorithm}

\section{Experimental Results}
In this section, we present numerical experiments on three nonlinear PDE control problems in the controlgym library \cite{zhang2023controlgym} that are, respectively, governed by the Burgers', Allen-Cahn, and Korteweg-de Vries equations. We introduce the setups of these environments below.

\noindent\textbf{(P1): Burgers' Equation} is a nonlinear PDE that models shock formation in water waves and gas dynamics. The temporal dynamics of the velocity $u(x, t)$ is
\begin{align*}
\frac{\partial u}{\partial t} + u\frac{\partial u}{\partial x} - \nu \frac{\partial^2 u}{\partial x^2} = a(x, t),
\end{align*}
where $\nu > 0$ denotes the viscosity coefficient, and the source term $a(x,t)$ is defined according to the forcing functions detailed in Section 2.2 of \cite{zhang2023controlgym}. In the experiments, we set $\nu = 10^{-4}$, the sampling time to $\Delta t = 0.05$, the integration time of controlgym's internal PDE solver to $dt=0.01$, the initial field to $u(x, t=0) = \alpha\cdot \mathrm{sech}(\frac{1}{\beta}(x - \frac{L}{2})$ with $\alpha \sim \texttt{uniform}(0.9, 1.1)$ and $\beta \sim \texttt{uniform}(0.04, 0.06)$, and $L=1$. The target field is set to $u_{r}(x) = 0.1\cdot \cos(2\pi x/L)$. 

We also set the problem time horizon to $300$, the number of states to $n_z = 128$, the number of actions to $n_a = 6$, the width of each distributed forcing support function to $0.15L$ (cf., \cite{zhang2023controlgym}), and the weighting matrices to $Q = R = I$. In DMDc, we set $p = 8$, $n_s = 4$, and use $a_k \sim \cN(0, 0.1\cdot I)$ to generate the exploratory system trajectory. 

\noindent\textbf{(P2): Allen-Cahn Equation} is a nonlinear PDE in materials science that models phase separation in binary alloy systems. The temporal dynamics of $u(x, t)$ in one spatial dimension, with $u = \pm 1$ indicating different phases, is described by
\begin{align*}
\frac{\partial u}{\partial t} - \nu^2 \frac{\partial^2 u}{\partial x^2} + V(u^3 - u) = a(x, t),
\end{align*}
where $\nu > 0$ is the diffusivity constant and $V$ is the potential constant. In the experiments, we set $\nu = 5\times 10^{-2}$, $V=5$, the sampling and integration times to $\Delta t = dt = 0.01$, the initial field to $u(x, t=0) = \alpha + (x - \frac{L}{2})^2 \cdot \cos(\frac{2\pi}{L}(x - \frac{L}{2}))$ with $\alpha \sim \texttt{uniform}(-0.1, 0.1)$, and $L=2$. The target field is set to $u_{r}(x) = -\cos(2\pi x/L)$.

We also choose the problem time horizon to be $80$, the number of states to $n_z = 256$, the number of actions to $n_a = 12$, the width of each distributed forcing support function to $0.05L$, and the weighting matrices to $Q = I$ and $R = 0.1\cdot I$. In DMDc, we set $p = 16$, $n_x = 8$, and use $u_t \sim \cN(0, 0.1\cdot I)$ to generate the exploratory trajectory.

\noindent\textbf{(P3): Korteweg-de Vries Equation} is a nonlinear PDE that models the propagation of solitary waves on shallow water surfaces.  The temporal dynamics of $u(x, t)$ is given by
\begin{align*}
\frac{\partial u}{\partial t} + \frac{\partial^3 u}{\partial x^3} - 6u\frac{\partial u}{\partial x} = a(x, t).
\end{align*}
We set the sampling time to $\Delta t = 0.01$, the integration time to $dt=0.001$, and the initial field to $u(x, t=0) = -\frac{\alpha}{2} \cdot \mathrm{sech}(\frac{\sqrt{\alpha}}{2}(x - \frac{L}{2}))$ with $\alpha \sim \texttt{uniform}(1, 3)$, and $L=20$. The target field is set to $u_{r}(x) = \sin(2\pi x/L)$.

We also choose the problem time horizon to be $200$, the number of states to $n_z = 256$, the number of actions to $n_a = 10$, the width of each distributed forcing support function to $0.05L$, and the weighting matrices to $Q = R = I$. In DMDc, we set $p = 16$ and $n_x = 8$, i.e., a 32-fold reduction in the state dimension, and use $u_t \sim \cN(0, 0.01\cdot I)$ to generate the exploratory trajectory.

\begin{figure*}[t]
	\centering
\includegraphics[width=0.94\textwidth]{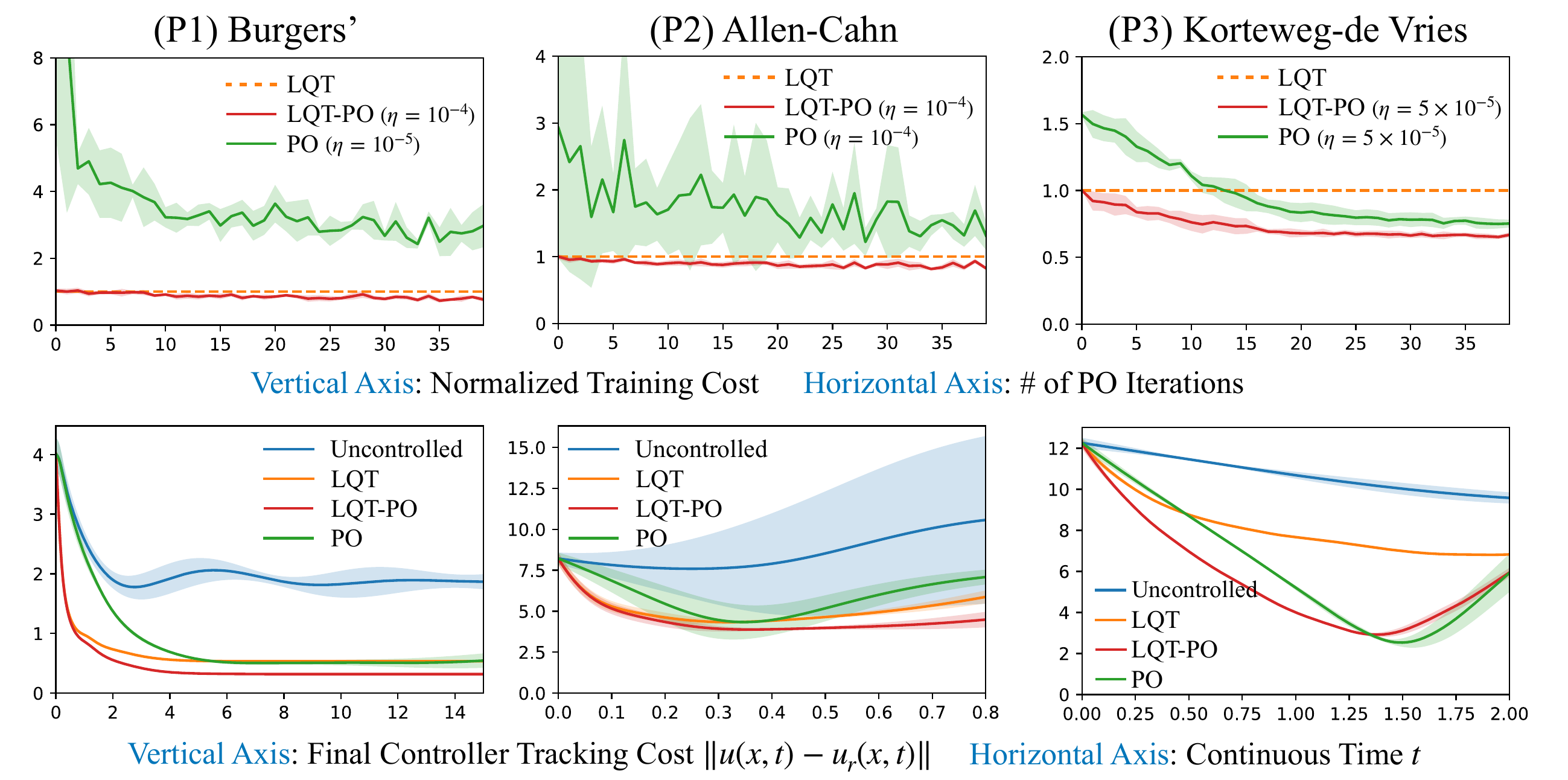}
 \vspace{-0.27em}
	\caption{\textit{Top}: Training curves of model-free PO for \textbf{(P1)}-\textbf{(P3)} averaged over $6$ random seeds with shaded regions denoting standard deviation. We normalize the vertical axis with respect to the cost of the LQT controller based on the DMDc model. For \textbf{(P1)}, the training process of PO (without a warm start) instantly destabilizes with $\eta=10^{-4}$. Hence, we select $\eta=10^{-5}$, the largest learning rate under which PO can consistently decrease the cost. \textit{Bottom}: The tracking costs of \textbf{(P1)}-\textbf{(P3)} with no control, LQT, LQT-PO after 40 iterations, and pure PO after 40 iterations. The costs are averaged over 10 trajectories with randomly sampled initial fields $u(x, t=0)$. The shaded region denotes the standard deviation.}
	\label{fig:training_curve}
 \vspace{-0.2em}
\end{figure*}

\noindent\textbf{PO parameters:} We choose the learning rates of LQT-PO as $\eta = 10^{-4}$ in \textbf{(P1)-(P2)} and $\eta = 5\times 10^{-5}$ in \textbf{(P3)}. The learning rates of pure PO is $\eta=10^{-5}$ in \textbf{(P1)}, $\eta=10^{-4}$ in \textbf{(P2)}, and $\eta = 5\times 10^{-5}$ in \textbf{(P3)}. See Figure \ref{fig:training_curve}. The smoothing radius of Algorithm \ref{alg:zeroth-order} is set to $r = 0.1$ in all cases.

Figure \ref{fig:training_curve} demonstrates that by adding a few iterations of PO, we can reduce the cost of the LQ tracking controller based on DMDc by $28.0\%$, $15.8\%$, and $36.4\%$, respectively, in the three PDE control tasks. Our results confirm the degradation of model-based controllers from the optimum in the presence of a large, unavoidable modeling gap. Consequently, there exist significantly outperforming controllers within the same reduced-order policy space, which could be found using PO. Furthermore, Figure \ref{fig:training_curve} shows that compared to PO from a zero initialization, PO with a (computationally cheap) warm start from the DMDc-based LQ tracking controller exhibits faster and more stable convergence across all three tasks. In the case of \textbf{(P1)}, the warm start also allows using a more aggressive learning rate in PO, leading to faster convergence without destabilizing the training process. 

\begin{figure}[t]
	\centering
\includegraphics[width=0.46\textwidth]{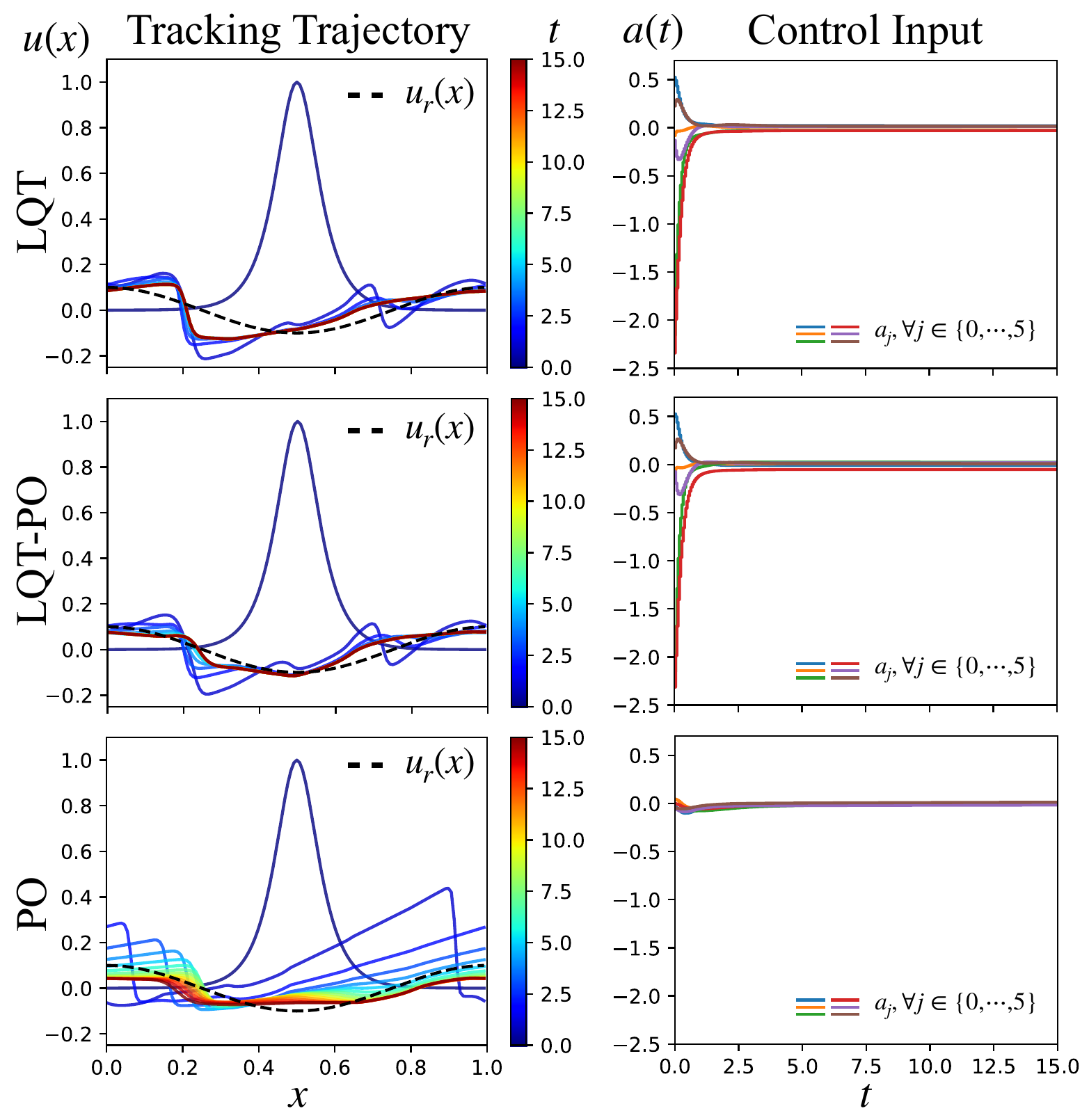}
  \vspace{-0.5em}
	\caption{\textbf{(P1):} Comparing model-based, PO with warm start, and pure PO control strategies with $40$ iterations of training budget for the latter two.}
	\label{fig:burgers}
  \vspace{-1em}
\end{figure}

\begin{figure}[t]
	\centering
	\includegraphics[width=0.46\textwidth]{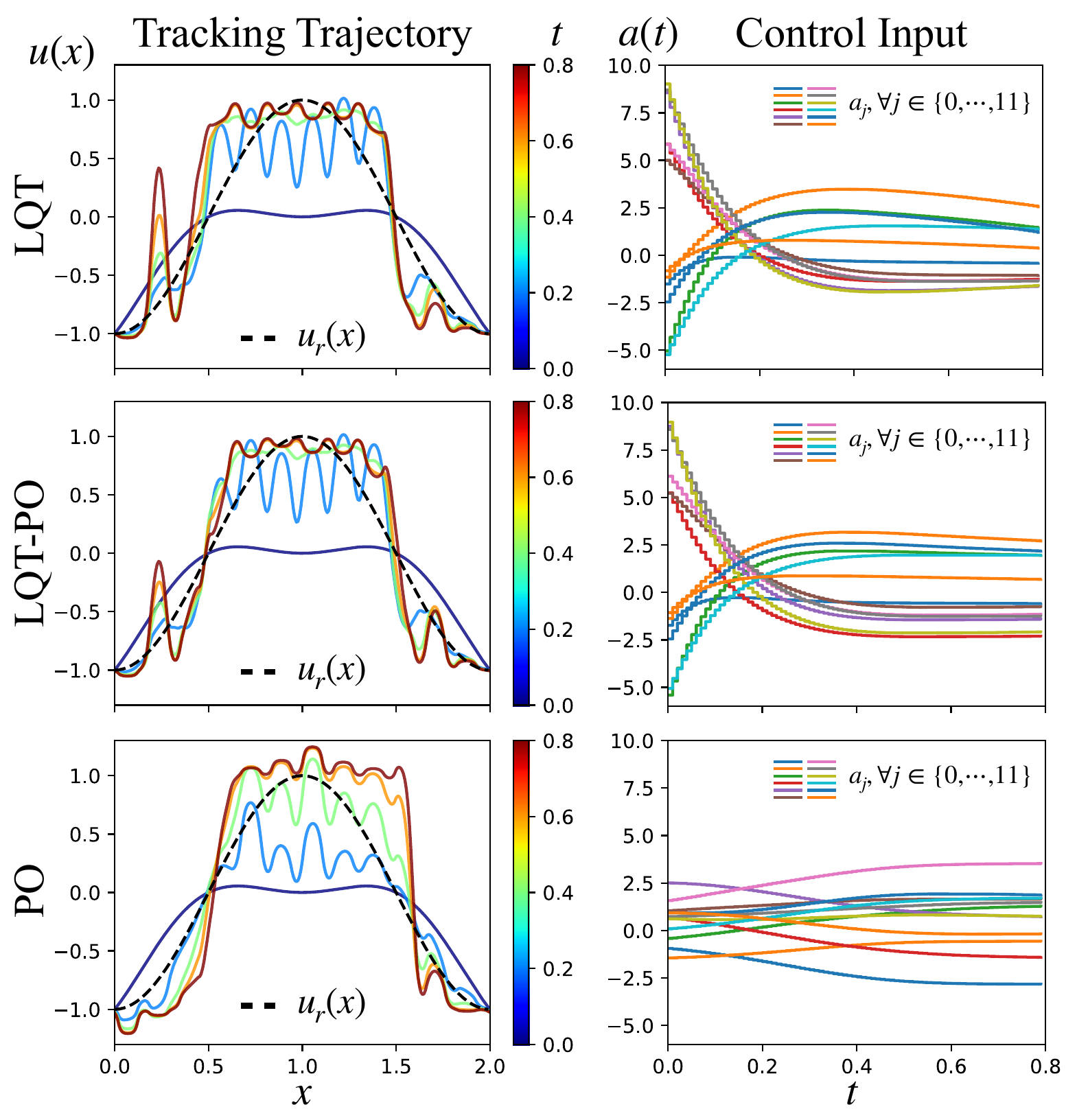}
  \vspace{-0.5em}
	\caption{\textbf{(P2):} Comparing model-based, PO with warm start, and pure PO control strategies with $40$ iterations of training budget for the latter two.}
	\label{fig:allencahn}
  \vspace{-1em}
\end{figure}

\begin{figure}[t]
	\centering
	\vspace{0.5em}\includegraphics[width=0.46\textwidth]{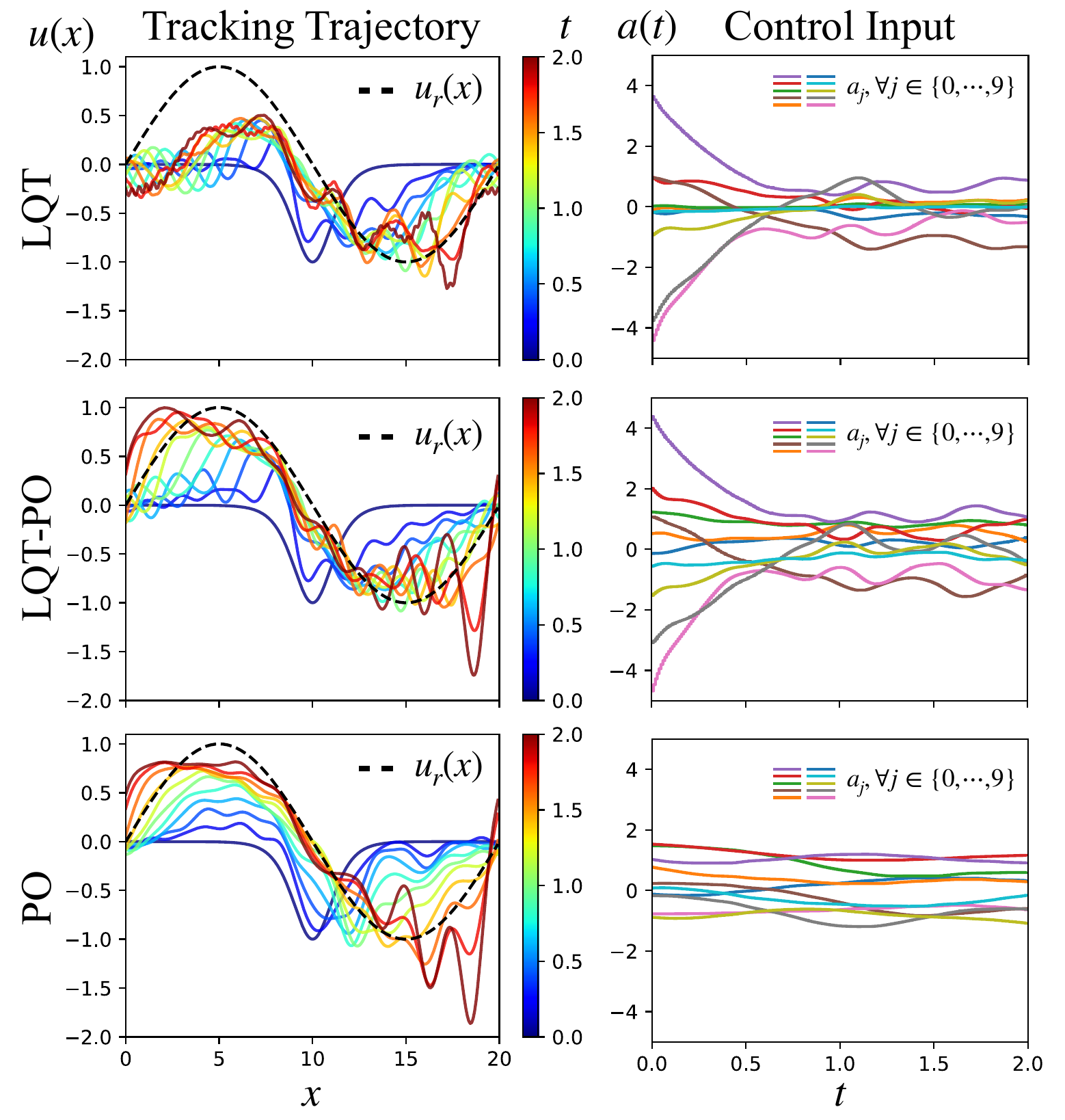}
  \vspace{-1em}
	\caption{\textbf{(P3):} Comparing model-based, PO with warm start, and pure PO control strategies with $40$ iterations of training budget for the latter two.}
	\label{fig:kdv}
  \vspace{-1em}
\end{figure}

In Figures \ref{fig:burgers}, \ref{fig:allencahn}, and \ref{fig:kdv}, we compare model-based, PO with warm start, and pure PO control strategies, where the latter two are subject to a certain budget of PO iterations. Each controller's behavior is evaluated with the initial field $u(x, t=0)$ set to the mean of their respective distributions. The results, as observed across Figures \ref{fig:burgers}-\ref{fig:kdv}, indicate that PO with a warm start achieves the best target state tracking among the three control strategies. The numerical results in this section confirm the effectiveness of our methodology.

\section{Conclusion}
We have augmented the ``reduce-then-design'' strategy with model-free PO for controlling spatio-temporal systems governed by nonlinear PDEs. Our numerical experiments demonstrate that PO significantly improves the performance of the ROM-based controller against the substantial modeling gap incurred from dimensionality reduction. Conversely, the ROM-based controller facilitates a warm start for PO, leading to accelerated and more stable learning than PO alone. An immediate future research topic is PDE control with imperfect state measurements, necessitating a data-driven state estimator in the controller design \cite{mowlavi2023may}. It is also worth considering nonlinear controller parametrizations, possibly through neural networks, and other PO schemes.


\small
\bibliographystyle{unsrt}
\bibliography{main}
\normalsize

\end{document}